\documentclass[preprint]{iucr}
\journalcode{J}

\usepackage{bm,amsmath,amssymb}

\begin{document}

\newcommand{\Tc}{T_{\text{c}}}
\newcommand{\Hcii}{H_{\text{c2}}}
\newcommand{\Hac}{H_{\text{ac}}}
\newcommand{\fms}{f_{\text{MS}}}
\newcommand{\MgB}{MgB$_2$}
\newcommand{\MgBC}{Mg(B$_{1-x}$C$_x$)$_2$}
\newcommand{\MgBCx}{Mg(B$_{0.93}$C$_{0.07}$)$_2$}
\newcommand{\MgMnB}{(Mg$_{1-x}$Mn$_x$)B$_2$}
\newcommand{\MgMnBx}{(Mg$_{0.995}$Mn$_{0.005}$)B$_2$}
\newcommand{\note}[1]{\textcolor{red}{#1}}
     
\title{Effects of magnetic and non-magnetic doping on the vortex lattice in {\MgB}}
%\shorttitle{Short Title} % for use in running heads (you will need to uncomment it).

\author[a]{E. R.}{Louden}
\author[b]{S.}{Manni}\aufn{Current address: Department of Physics, Indian Institute of Technology Palakkad, Kerala, India}
\author[a]{J. E.}{Van Zandt}
\author[a]{A. W. D.}{Leishman}
\author[b,c]{V.}{Taufour}\aufn{Current address: Department of Physics and Astronomy, University of California Davis, 95616 Davis, CA, USA}
\author[b,c]{S. L.}{Bud'ko}
\author[d]{L.}{DeBeer-Schmitt}
\author[e]{D.}{Honecker}
\author[e]{C. D.}{Dewhurst}
\author[b,d]{P. C.}{Canfield}
\cauthor[a]{M. R.}{Eskildsen}{eskildsen@nd.edu}{}

\aff[a]{Department of Physics, University of Notre Dame, 46556 Notre Dame, IN, \country{USA}}
\aff[b]{Department of Physics and Astronomy, Iowa State University, 50011 Ames, IA, \country{USA}}
\aff[c]{Division of Materials Science and Engineering, Ames Laboratory, 50011 Ames, IA, \country{USA}}
\aff[d]{Large Scale Structures Group, Neutron Sciences Directorate, Oak Ridge National Laboratory, 37831 Oak Ridge, TN, \country{USA}}
\aff[e]{Institut Laue-Langevin, 71 avenue des Martyrs, CS 20156, F-38042 Grenoble cedex 9, \country{France}}

%Use \shortauthor to indicate an abbreviated author list for use in running heads.
%\shortauthor{Soape, Author and Doe}


\maketitle

\begin{synopsis}
The vortex lattice phase diagram, order and transition kinetics in superconducting {\MgB} doped with either manganese or carbon was studied using small-angle neutron scattering.
\end{synopsis}

\newpage
\begin{abstract}
Using small-angle neutron scattering we have studied the vortex lattice in superconducting {\MgB}, doped with either manganese or carbon to achieve a similar suppression of the critical temperature.
Measurements were performed with the magnetic field applied along the $c$-axis, where the vortex lattice in pure {\MgB} is known to undergo a field-and temperature-driven 30$^{\circ}$ rotation transition.
For Mn-doping, the vortex lattice phase diagram remains qualitatively similar to that of pure {\MgB}, indicating only a modest effect on the vortex-vortex interaction.
In contrast, the vortex lattice rotation transition is completely suppressed in the C-doped case, likely due to a change in the electronic structure which affects the two-band/two-gap nature of superconductivity in {\MgB}.
The vortex lattice longitudinal correlation length shows the opposite behavior, remaining roughly unchanged between pure and C-doped {\MgB} while it is significantly reduced in the Mn-doped case.
However, the extensive vortex lattice metastability and related activated behavior, observed in conjunction with the vortex lattice transition in pure {\MgB}, is also seen in the Mn doped sample.
This shows that the vortex lattice disordering is not associated with a substantially increased vortex pinning.
\end{abstract}

\newpage
%%% INTRODUCTION
\section{Introduction}
Vortex matter in type-II superconductors is highly sensitive to the environment provided by the host material.
As an example one can consider the vortex lattice (VL) symmetry, orientation and even transitions between different orientations, which is governed by anisotropies in the screening current plane perpendicular to the applied field and the associated nonlocal vortex-vortex interactions~\cite{Kogan:1997vm,Kogan:1997wy,Franz:1997jn,Agterberg:1998wo,Ichioka:1999aa}.
As no material is perfectly isotropic, the observation of anisotropy effects in superconductors is near ubiquitous, with examples given in our recent review on magnetic small-angle neutron scattering (SANS)~\cite{Muhlbauer:2019jt}.
The VL may also be used to probe the superconducting state itself, as demonstrated by the recent evidence for broken time-reversal symmetry in the topological superconductor UPt$_3$\cite{Avers:2020wx}.
Here we focus on the how the VL in {\MgB} is affected by both non-magnetic and magnetic doping.

The VL phase diagram in {\MgB} with $\bm{H} \parallel \bm{c}$ consists of three different VL phases, shown later in Fig.~\ref{VLPhaseDiagram}(a). All the VL phases have a triangular symmetry, differing only in their orientation relative to the crystalline axes ~\cite{Cubitt:2003ip,Das:2012cf}.
In the F and I phases the VL is oriented along the crystalline $\bm{a}$- and $\bm{a}^*$-axis directions respectively.
In the intermediate L phase the VL undergoes a continuous $30^{\circ}$ rotation transition between these two high-symmetry directions within the hexagonal crystalline basal plane.
The VL rotation is due to the field- and temperature dependent supercarrier density within the $\pi$- and $\sigma$-bands in {\MgB}, resulting in a change of the relative strength of their competing anisotropies~\cite{Zhitomirsky:2004jq,Hirano:2013jx,Olszewski:2020jy}.
The VL phase diagram can be tuned by rotating the applied field away from the $c$-axis, which first suppresses and eventually eliminates the intermediate L phase, and leads to a discontinuous transition between the F and I phases~\cite{Leishman:2021ii}.

In addition to a rich phase diagram the VL in {\MgB} displays extensive metatstability upon cooling or heating across the equilibrium phase boundaries, attributed to the presence of VL domain boundaries~\cite{Das:2012cf,Rastovski:2013ff}.
Recent studies of the transition kinetics, using vortex motion to gradually drive a supercooled VL from the metastable to the equilibrium state, allowed a determination of the associated activation barrier~\cite{Louden:2019bq,Louden:2019wx}.
Here, the activation barrier was found to increase as the metastable state is suppressed, corresponding to an aging of the VL.

To further explore how the vortex-vortex interaction in {\MgB} can be modified we have used SANS to study samples doped with either magnetic manganese (Mn) or non-magnetic carbon (C) atoms on the Mg and B sites respectively.
Here, doping levels were chosen to achieve roughly the same suppression of the critical temperature ($\Tc$).
In the Mn-doped case the VL phase diagram remained largely the same as that for pure {\MgB}, although a clear disordering was observed through a decrease in the longitudinal VL correlation length.
Conversely, the C-doped sample demonstrated a complete suppression of the L phase compared to pure {\MgB}, while the correlation length remained unaffected.
Finally, VL metastability and activated behavior was still observed in the Mn-doped sample, although with a slightly higher activation barrier compared to pure {\MgB}.

%%% Crystal Growth and Characterization
\section{Crystal Growth and Characterization}
Single crystals of {\MgBC} and {\MgMnB} were grown by a high pressure flux method from excess solution of magnesium~{\cite{Karpinski:2003aa,Mou:2016jl,Taufour:2021aa}.
Elemental Mg, B, and dopant Mn or C were packed inside a 1~cm diameter boron nitride (BN) crucible in the ratio of Mg:B:(Mn/C) = 1:0.7:$x$.
The dopant C/Mn was placed at the bottom of the crucibles for maximum incorporation within the Mg melt, and above it a pressed Mg-pellet and B powder were placed.
To reduce neutron absorption during the SANS experiments isotopically pure $^{11}$B was used for the growth.
However, due to contamination from the BN crucibles made with natural boron the {\MgB} crystals contains $\sim 5$\% neutron absorbing $^{10}$B~\cite{Cubitt:2003ip}.
The crucible was tightly packed with BN powder to prevent any Mg leakage at high pressure and temperature.
Using a cubic anvil furnace, a pressure of 3.5~GPa was applied to the crucible at room temperature, and it was then heated to 1430$^{\circ}$C and held at this temperature for 1~hr.
Afterwards the temperature was decreased to the 650$^{\circ}$C melting point of Mg over a time of 6~hrs.
After completion of the crystal growth, the furnace heating was turned off to achieve a faster cooling and the pressure was released.
This procedure yields MgB$_2$ single crystals embedded inside the solidified Mg-melt.
Individual single crystals were extracted by distilling the excess Mg at 750 $^{\circ}$C inside an evacuated quartz tube.
The crystals for the SANS experiments had a flat-plate morphology, with side lengths of roughly half a millimeter and masses in the 50--100~$\mu$g range.

For the SANS measurements, C- and Mn-doped crystals were chosen to have roughly equal $\Tc$, determined from zero field cooled magnetization measurements.
For the selected {\MgBC} crystals $\Tc \approx 34$~K, indicating a dopant concentration of $x = 0.04 - 0.07$~\cite{Wilke:2004kj,Kazakov:2005ip,Wilke:2007fb}.
For the {\MgMnB} crystals $\Tc \approx 33$~K, corresponding to a doping of $x = 0.005 \pm 0.001$~\cite{Rogacki:2006ei}.
Field-temperature phase diagrams for both the carbon and the manganese doped samples are shown in Fig.~\ref{Hc2}.
Linear fits to the upper critical field ($\Hcii(T)$) near the $\Tc$ for $\bm{H} \parallel \bm{c}$ yields slopes of $d\Hcii/dT = -0.22$~T/K and $-0.090$~T/K for the C- and Mn-doped samples respectively.
Also shown for comparison is $d\Hcii/dT = -0.11$~T/K for pure Mg$^{11}$B$_2$~\cite{Das:2012cf}.
Consistent with earlier reports, the slope for {\MgMnBx} is roughly the same as for undoped {\MgB}~\cite{Rogacki:2006ei}, while it doubled for {\MgBCx}~\cite{Kazakov:2005ip,Angst:2005gk}.

% The zero-temperature value of the upper critical field is estimated using the Werthamer-Helfand-Hohenberg expression 
% \begin{equation}
%     \Hcii(0) \approx 0.7 \, \Tc \left. \left( \frac{\text{d} \Hcii}{\text{d} \Tc} \right) \right|_{\Tc},
% \end{equation}
% using the derivative $\text{d}\Hcii/\text{d}\Tc$ near {$\Tc$} obtained from the data in Fig.~{\ref{Hc2}}.
% For the carbon doped sample this yields $\Hcii(0) \sim 5$~T, in good agreement with that reported by Amvrazi and Pissas for $x = 0.04 - 0.1$~\cite{Amvrazi:2014jp}, but smaller than than Kazakov {\em et al.} who found $\Hcii$ closer to 8~T for $x = 0.05 - 0.095$~\cite{Kazakov:2005ip}.
% For the manganese doped sample the extrapolation yields $\Hcii(0) \sim 2$~T, comparable to the value of $\sim 2.5$~T for $x = 0.0042$ reported by Rogacki {\em et al.}~\cite{Rogacki:2006ei}.
    
% C-doped: Kazakov x = 4% gives Tc = 34.3 K
% Mn-doped: Rogacki x = 0.5% gives Tc = 33 K
% The crystals were grown using isotopically enriched $^{11}B$ to reduce neutron absorption. The critical temperature of the {\MgBC} sample was determined from magnetization measurements to be 33~K. Through extrapolation to $T~=~0$, the critical magnetic field was estimated to be $\sim 7$~T. From these critical values, the dopant concentration was estimated to be $x = 0.07 \pm 0.1$.  Similarly for the {\MgMnB} sample, the critical values were determined to be \Tc $= 30$~K and \Hc $= 2$~T.

%%% SANS
\section{Small-Angle Neutron Scattering}
The SANS measurements were performed using a conventional geometry for VL studies with the applied field parallel to the incident neutron beam~\cite{Muhlbauer:2019jt}.
Experiments were carried out at the D33 beam line at the Institut Laue Langevin (ILL)~\cite{5-42-420},
the GP-SANS beam line at Oak Ridge National Laboratory (ORNL)~\cite{Heller:2018},
and the NG7 beam line at the NIST Center for Neutron Research (NCNR).
The crystal orientations was verified using a back-scattering X-ray Laue camera system (Multiwire Laboratories Ltd) at the ORNL Spallation Neutron Source X-ray laboratory.

The magnetic field was applied parallel to the crystalline $c$-axis, and the sample and magnet were rotated together to satisfy the Bragg condition for the VL with scattering angles given by
$\sin \theta_0 \approx \theta_0 = Q_{\text{VL}} \lambda_n /4 \pi$~\cite{Muhlbauer:2019jt}.
Here $Q_{\text{VL}} = 2\pi \sqrt{2B/\sqrt{3}\Phi_0}$ is the magnitude of the scattering vector for a triangular VL, $\Phi_0 = h/2e = 2068$~T~nm$^2$ is the superconducting flux quantum, and $\lambda_n$ is the neutron wavelength.
% NCNR 09/2016: 0.7~nm, ILL 09/2016: 0.7~nm.}
The magnetic induction $B$ is equal to the applied field $\mu_0 H$ as the magnetization for our samples is negligible at the fields used in the SANS experiments.
Diffracted neutrons were collected on a 2D position sensitive detector.
The orientation of the in-plane crystaline $\bm{a}$ and $\bm{a}^*$ axes were determined by X-ray Laue diffraction.

The SANS measurements were performed at temperatures between $T = 2$ and $20$~K and magnetic fields in the range $\mu_0 H = 0.3$ to $2.5$~T. 
For the determination of the equilibrium phase diagram the VL was prepared at each field and temperature by applying a damped field oscillation, which is known to produce well-ordered equilibrium VL states in pure {\MgB}~\cite{Rastovski:2013ff}.
Studies of non-equilibrium VLs and the metastable-to-equilibrium transition kinetics were preceded by cooling across the F-L phase transition in a constant field to render the VL in a metastable state~\cite{Das:2012cf}.
Subsequently, the VL was gradually driven from the metastable F phase to the equilibrium L phase by successive application of ac magnetic field oscillations with an amplitude of $0.93$~mT and a frequency of 250~Hz~\cite{Louden:2019bq,Louden:2019wx}.
Here, the ac field is parallel to the dc field.

%%% RESULTS
\section{Results}
The results of our SANS measurements related to the equilibrium VL phase diagram, the structural ordering of the VL, and the metastable-to-equilibrium transition kinetics are presented in the following.

% Equilibrium VL phase diagram
\subsection{Equilibrium VL phase diagram}
As already discussed, the equilibrium VL phase diagram in pure {\MgB}, shown in Fig.~\ref{VLPhaseDiagram}(a)(see below), consists of three different triangular  phases~\cite{Das:2012cf}: an F phase with Bragg peaks along the crystalline $\bm{a}$-axis direction, an I phase with peaks along the $\bm{a}^*$-axis direction, and an intermediate L phase where the VL undergoes a continuous $30^{\circ}$ rotation transition.
In the F and I phases, where the VL is oriented along one of the high-symmetry crystalline directions, SANS diffraction pattern show six Bragg peaks.
In contrast, in the L phase, there is a two-fold degeneracy with respect to the direction of the VL rotation (clockwise vs counter-clockwise) leading to the presence of oppositely rotated domains and six pairs of azimuthally split Bragg peaks.
Figure~{\ref{DifPats}} shows representative SANS diffraction patterns obtained in {\MgBCx} and {\MgMnBx}.
To conserve beam time, not all VL peaks were rocked through the Bragg condition; however their positions can be determined by successive $60^{\circ}$ rotations around the origin of reciprocal space ($Q = 0$) as indicated by the circles.
In the case of carbon doping a F phase VL, shown in Fig.~\ref{DifPats}(a), is observed at all the fields and temperatures where it was possible to image the VL by SANS.
In comparison, all three VL phases are observed for the manganese doped sample as illustrated in Figs.~\ref{DifPats}(b) - \ref{DifPats}(d).

In undoped {\MgB} thermal fluctuations are insufficient to overcome the activation barrier required for transitions between the F, L and I phases, giving rise to extensive metastability~\cite{Louden:2019bq}.
However, an equilibrium VL configuration may be achieved at any given field and temperature by introducing vortex motion, for example through the application of a damped field oscillation~\cite{Das:2012cf}.
This is also the case for {\MgMnBx}, as evident from the comparison between different field and temperature histories in Fig.~\ref{MnDopedFieldHist}, showing the VL scattered intensity versus the azimuthal angle ($\varphi$) for one $60^{\circ}$ segment of the diffraction pattern.
Here, the reference direction $\varphi_0$ is along the $\bm{a}$-axis, such that a single peak at $\varphi = \varphi_0$ corresponds to the F phase whereas split peaks around $\varphi_0$ correspond to the L phase.
As seen in Fig.~\ref{MnDopedFieldHist}(a), field cooling from $T > \Tc$ results in a majority F phase VL.
The smaller shoulders on the main peaks indicates the presence of minor L phase domains.
In contrast, an L phase VL is obtained following either a zero field cooling and a subsequent field application at base temperature as shown in Fig.~\ref{MnDopedFieldHist}(b), or by cooling in a high field followed by a field reduction as shown in Fig.~\ref{MnDopedFieldHist}(c).
In both cases significant vortex motion was induced by the field change.
Applying a damped field oscillation, with an initial amplitude of 75~mT, causes a transition from the VL F phase in Fig.~\ref{MnDopedFieldHist}(a) to an L phase shown in Fig.~\ref{MnDopedFieldHist}(d).
This confirms that the F phase is metastable, while the L phase is the equilibrium VL configuration at the specific field and temperature.
This is consistent with the results in Figs.~\ref{MnDopedFieldHist}(e) and \ref{MnDopedFieldHist}(f).
Here, the additional vortex motion induced by the damped field oscillation only gives rise to slight changes in the population of the two domain orientations as indicated by the relative intensity of the peaks, while the VL remains in the L phase with an unchanged peak splitting $\Delta \varphi \approx 22^{\circ}$.

The equilibrium VL phase diagram are summarized in Figs.~\ref{VLPhaseDiagram}(b) and \ref{VLPhaseDiagram}(c) for {\MgBCx} and {\MgMnBx} respectively, with that for undoped {\MgB} shown in Fig.~\ref{VLPhaseDiagram}(a) for comparison.
At each field and temperature measured, a damped field oscillation with an initial amplitude of 50~mT was applied before imaging the VL to ensure an equilibrium configuration.
Symbols indicate where SANS imaging of the VL was performed, with the limiting factor being the vanishing scattering intensity at both high temperatures and fields.
In particular, even a modest extension along the field axis would require excessive count times due to the exponential decrease of the intensity.
For the carbon-doped sample only the F phase is observed, as seen in Fig.~\ref{VLPhaseDiagram}(b), although other VL phases may exist at high fields and/or temperatures.
That said, the measurements extended up to $2.5$~T$ \sim 0.44 \Hcii$, which is an order of magnitude higher than the F-L transition observed in undoped {\MgB}, both on an absolute (0.2~T) as well as a relative scale ($ \sim 0.06 \Hcii$).
The VL phase diagram for {\MgBCx} thus appears to be qualitatively different from that of pure {\MgB}.
In comparison, the phase diagram for {\MgMnBx}, shown in Fig.~\ref{VLPhaseDiagram}(c), only differs quantitatively from that of undoped {\MgB}.
Specifically, the F-L phase boundary has moved to higher fields and the L-I boundary shifted to lower temperatures resulting in a shrinking of the L phase.
As seen in Fig.~\ref{VLPhaseDiagram}(d) the VL peak splitting angle $\Delta \varphi$, determined from multi-peak Gaussian fits to the azimuthal intensity distribution, saturates at the highest measured fields.
This suggests that the L phase in {\MgMnBx} extends to the upper critical field at low temperature, similar to the situation in undoped {\MgB}.
Still, the low temperature saturation angle $\Delta \varphi \approx 56^{\circ}$ is closer to the I phase with ($60^{\circ}$) than the value of $\approx 44^{\circ}$ for the pure compound.

% Longitudinal correlation length
\subsection{Longitudinal Correlation Length}
Both lattice imperfections and the finite experimental resolution causes the VL reflections to broaden in reciprocal space, with scattering occurring for a range of angles around the Bragg condition. 
Figure~{\ref{RockingCurves}} shows the scattered intensity as the VL is rotated through the Bragg condition at $\theta = \theta_0$ in a typical ``rocking curve'' for for both pure and doped {\MgB}.
While the rocking curve for {\MgBCx} only show a slight broadening compared to the undoped compound, that of {\MgMnBx} is substantially wider.

Spatial correlations for the VL are expected to decay exponentially with distance, which, in the absence of experimental broadening, lead to rocking curves with a Lorentzian line shape.
This allows a determination of the longitudinal correlation length (i.e. along the field direction),
\begin{equation}
    \label{CorLength}
    \zeta_{\text{L}} = \frac{d_0}{\pi w_L},
\end{equation}
where $d_0 = \sqrt{2\Phi_0/\sqrt{3}B}$ is the vortex separation (69~nm at 0.5~T), and $w_L$ is the Lorentzian full width at half maximum (FWHM) in radians.
In situations where the intrinsic broadening is comparable to the instrumental resolution, rocking curves are best described by a Voigt profile:
\begin{equation}
    \label{Voigt}
    V(\theta) = \int_{-\infty}^{\infty} G(\theta') \; L(\theta - \theta') \; d\theta'.
\end{equation}
This is a convolution of the Lorentzian ($L$) representing the intrinsic width and a Gaussian ($G$) determined by the experimental resolution~\cite{Eskildsen:1998aa,Louden:2019jb}.
Curves in Fig.~\ref{RockingCurves} are Voigt fits to the data.
In all cases the Gaussian width, estimated from the neutron wavelength, wavelength spread and beam collimation used in the respective SANS experiment~\cite{Eskildsen:1998aa}, was kept fixed during the fitting.
Compared to the rather poor experimental resolution within the detector plane evident from Fig.~\ref{MnDopedFieldHist}, the perpendicular Gaussian width is reduced by a factor $(Q_{\text{VL}} \lambda_n)/2\pi \sim 10^{-2}$~\cite{Muhlbauer:2019jt}.
For the present measurements the Lorentzian width dominated the experimental resolution, allowing a precise determination of $\omega_L$ and thus the longitudinal correlation length.
From the Lorentzian widths, listed in each panel of Fig.~\ref{RockingCurves}, we find $\zeta_{\text{L}} \sim 9$~$\mu$m for undoped {\MgB}, $\zeta_{\text{L}} \sim 7$~$\mu$m for {\MgBCx}, and $\zeta_L \sim 2$~$\mu$m for {\MgMnBx}.
One should note that while the rocking curves for the doped samples were measured during the same SANS experiment, and are thus directly comparable, the one for the pure compound was obtained separately.
A comparison between {\MgBCx} and {\MgB} may therefore be affected by systematic errors, and their correlation lengths should be considered equivalent.

% MS to ES transition kinetics
\subsection{Metastable to Equilibrium State Transition Kinetics}
The presence of robust metastable states makes it possible to study the VL transition kinetics and determine the activation barrier separating the metastable and equilibrium states in {\MgMnBx}, using the same stop-motion technique as for undoped {\MgB}~\cite{Louden:2019bq,Louden:2019wx}.
First, a pristine metastable F phase VL is prepared by applying a damped field oscillation at 0.5~T and 25~K and then cooling the sample across the F-L phase boundary to a final temperature of 2~K.
A measurement sequence is then performed, alternating between SANS imaging of the VL and applications of small-amplitude ac magnetic field cycles.
The latter introduces a modest amount of vortex motion which gradually drives the VL towards the equilibrium state.

The results of the measurement sequence are summarized in Fig.~\ref{MnDopedKineticsActivation}.
Panel (a) shows an example of the the azimuthal intensity distribution, with scattering from both metastable F phase (MS) and equilibrium L phase (ES1/ES2) domains clearly evident.
The population of each domain orientation is proportional to the corresponding peak intensity, determined from a three Gaussian fit, and the volume fraction remaining in the metastable state is thus given by
\begin{equation}
    \label{fMSdef}
    \fms = \frac{I_{\text{MS}}}{I_{\text{ES1}} + I_{\text{MS}} + I_{\text{ES2}}}.
\end{equation}
Here, the width used for the Gaussian fits is determined from the pristine metastable F phase (0 cycles) and then kept fixed~\cite{Louden:2019bq}.
The evolution of $f_{\text{MS}}$ as a function of the cumulative ac field cycle count ($n$) is shown in Fig.~\ref{MnDopedKineticsActivation}(b).

Assuming an activated behavior, with the ac field amplitude (${\Hac}$) and $n$ taking the role of respectively an effective ``temperature'' and ``time'', the evolution of the metastable volume fraction is given by~\cite{Louden:2019bq}
\begin{equation}
    \label{fMSrate}
    \frac{d \fms}{dn} = - \fms \exp[ -\tilde{H}/\Hac].
\end{equation}
The activation field $\tilde{H}$, representing the barrier between metastable and equilibrium VL domain orientations, is determined experimentally from adjacent values of $\fms(n)$ by 
\begin{equation}
    \label{Hact}
    \tilde{H} = - \Hac \ln \left[ - \frac{\ln \{ \fms(n_{i+1})/\fms(n_i) \}}{n_{i+1} - n_i} \right]
\end{equation}
and shown in Fig.~\ref{MnDopedKineticsActivation}(c).
The increasing $\tilde{H}(n)$ is equivalent to an aging of the VL~\cite{Henkel:2010wn}.

Similar to pure {\MgB}, the metastable volume fraction for the Mn-doped sample in Fig.~\ref{MnDopedKineticsActivation}(b) is well fitted by~\cite{Louden:2019bq}
\begin{equation}
  \fms(n) = \exp \left[ \alpha + \beta (n+1)^{\gamma} \right].
  \label{fMSfit}
\end{equation}
Here $\beta \approx - \alpha$ is introduced to accommodate a value of $\fms(0)$ slightly less than one.
From the fitted values of $\alpha$ and $\gamma$ one obtains two physical quantities~\cite{Louden:2019bq}.
First, $\alpha_{\text{Mn}} \sim \tfrac{1}{3} \alpha_{\text{pure}}$ yields a roughly one order of magnitude increase in the saturation value $\fms (n \rightarrow \infty) = e^{\alpha}$ (vertical asymptote of the fits in Fig.~\ref{MnDopedKineticsActivation}(c).
Second, $\alpha$ and $\beta$ combined allows a calculation of the initial value of the activation barrier $\tilde{H}(\fms = 1)$, which remains largely unchanged compared to pure {\MgB}.

%%%%% DISCUSSION %%%%%    
\section{Discussion}
% Hirano, VL phase diagram, pi- and sigma-band anisotropy
The SANS studies of the VL phase diagrams in doped {\MgB} furthers our understanding of the superconducting properties of this material.
Theoretically, it was found that the stability of the VL F and I phases may be attributed to the Fermi surface anisotropy of the $\pi$- and $\sigma$-bands respectively~\cite{Hirano:2013jx}.
Combined with the suppression of $\pi$-band superconductivity by an applied field~\cite{Eskildsen:2002ih}, this leads to the observed VL phase diagram~\cite{Cubitt:2003ip,Das:2012cf}.
The $F \rightarrow L \rightarrow I \rightarrow L$ phase sequence along $\Hcii$ from $\Tc$ to zero temperature remains a robust feature upon variations of both the intra- and interband pairing constants, as long as none of them vanish~\cite{Hirano:2013jx}.
However, the relative fraction of the phase diagram occupied by the different states will change depending on the gap ratio $\Delta_{\pi}/\Delta_{\sigma}$.
Setting both the $\sigma$-band pairing and the interband coupling equal to zero completely eliminates the I phase and substantially reduces the range of stability of the L phase.
This picture, combined with results from previous experimental studies of {\MgB} doped with either carbon or manganese, is consistent with our SANS results, as shall be discussed in the following.

Between the two dopants studied, Mn affects the superconductivity in {\MgB} in the simplest manner.
 Within the explored doping range, $\Tc$ and $\Hcii(0)$ for $\bm{H} \parallel \bm{c}$ are both suppressed while maintaining a linear relationship, and $d\Hcii /dT$ at the critical temperature remains unaffected~\cite{Rogacki:2006ei}.
This implies minimal changes in the intraband scattering for both the $\pi$- and $\sigma$-bands.
Furthermore, point contact spectroscopy shows that the distinct superconducting gaps for the $\pi$- and $\sigma$-bands are preserved up to the highest achievable doping levels ($\Tc = 10$~K), and interband scattering must therefore remain weak~\cite{Gonnelli:2006hq}.
The conclusion of these studies is that the Mn-doping predominantly influences the superconducting properties through spin-flip scattering, with the reduction of $\Tc$ being due to magnetic pair-breaking.
This is consistent with our observation of a VL phase diagram which is qualitatively similar to that of pure {\MgB}.
At the quantitative level, the gap ratio $\Delta_{\pi}/\Delta_{\sigma}$ increases slightly with Mn-doping~\cite{Gonnelli:2006hq}, which is predicted to increase the range of stability of the F and L phases and decrease the I phase~\cite{Hirano:2013jx}.
In comparison we observe an increase of both the F and I phases at the expense of the L phase, as seen in Fig.~\ref{VLPhaseDiagram}(a) and \ref{VLPhaseDiagram}(c).
This suggests that either the simple model of Hirano {\em et al.} does not fully capture the behavior of the VL in {\MgB}, or that the slightly increased intraband scattering due to the Mn doping changes the effective range of the nonlocal vortex-vortex interaction and slightly shifts the phase boundaries.
We will return to the second possibility later.

% C-doped VL phase diagram
In the case of C-doping the effect on the superconducting properties is more complex~\cite{Gurevich:2003ei,Wilke:2004kj,Kortus:2005jf,Samuely:2005ka,Kortus:2005jy}.
As was the case for Mn-doping, point contact spectroscopy studies found a modest reduction of $\Delta_{\sigma}$ while $\Delta_{\pi}$ remained largely unaffected for C-doping concentrations comparable to that studies by SANS~\cite{Gonnelli:2005jz,Samuely:2005dv,Szabo:2007dm}.
In the absence of other effects one would therefore expect only minor modifications of the VL phase diagram, in analogy with the Mn-doped case.
Therefore the complete suppression of the L and I phases thus clearly demonstrate the presence of additional effects.
Theoretically, the lack of a doping dependence for $\Delta_{\pi}$ and $\Delta_{\sigma}$, the reduction of $\Tc$, and increase of both $\Hcii(0)$ and $d\Hcii /dT$ %~\cite{Wilke:2004kj,Kazakov:2005ip}
is explained by compensating effects of increased intra- and interband scattering and a reduction of $\sigma$-band density of states due to the electron doping by carbon~\cite{Kortus:2005jf,Samuely:2005ka,Kortus:2005jy,Angst:2005gk,Kazakov:2005ip,Wilke:2007fb}.
Experimentally, this is also found to increase the resilience of the $\pi$-band superconductivity to magnetic fields~\cite{Szabo:2007dm}.
Both the reduction of the $\sigma$-band density of states and the increased robustness of the $\pi$-band superconductivity will favor the VL F phase at the expense of the L and I phases~\cite{Hirano:2013jx}, consistent with our SANS results.

% Nonlocality
We now return to the discussion of potential modifications to the VL phase diagram due to a reduction of the mean free path, and by extension the range of the nonlocal vortex-vortex interactions.
To put this in context, consider the triangular to square VL symmetry transition observed in most superconductors with a four-fold basal plane anisotropy~\cite{Muhlbauer:2019jt}.
This transition is driven solely by a change of the vortex density, and occurs when the separation is a certain fraction of the nonlocality range~\cite{Kogan:1997vm,Kogan:1997wy}.
As shown for Lu(Ni$_{1-x}$Co$_x$)$_2$B$_2$C the symmetry transition is shifted to progressively higher fields, as the mean free path is reduced with increasing Co-doping~\cite{Gammel:1999aa}.
In contrast, the rotation between the F and I phases in {\MgB} is driven by the suppression of the $\pi$-band superconductivity with increasing field, rather than the accompanying change of the vortex density.
While the slight change of the VL phase diagram in Mn-doped {\MgB} may be due to a reduced mean free path, it is unlikely to explain the complete suppression of the L and I phase in the C-doped case.
Rather, one would expect the opposite effect as C-doping has been found to enhance the $\pi$-band intraband scattering more rapidly than that of the $\sigma$-band~\cite{Szabo:2007dm}, which will suppress the influence of the $\pi$-band Fermi surface anisotropy and thus disfavor the VL F phase.

% Disorder, lack of strong pinning, transition kinetics
Finally, we comment on the structural properties of the VL together with the metastability and transition kinetics in Mn-doped {\MgB}.
While the longitudinal correlation length is considerably decreased when compared to the pure and C-doped samples, indicated by the broadening rocking curves seen in Fig.~\ref{RockingCurves}(c), clear VL Bragg peaks are still observed on the SANS detector as shown in Figs.~\ref{DifPats}(b)-\ref{DifPats}(d).
That said, some in-plane broadening is observed relative to the C-doped sample, shown in Fig.~\ref{DifPats}(a).
However, due to the poor in-plane resolution a quantitative comparison is not feasible as the peaks widths for both pure and C-doped {\MgB} are resolution limited.
We note that a substantial rocking curve broadening is observed other in materials that include magnetic elements such as ErNi$_2$B$_2$C~\cite{Yaron:1996aa} and TmNi$_2$B$_2$C~\cite{Eskildsen:1998aa}, regardless of whether the magnetic moments order.
At this time this is simply an observation for which we do not have a clear explanation.
The decreased correlation length in the Mn-doped sample is likely not indicative of substantially increased vortex pinning.
This notion is supported by the azimuthal widths in Fig.~\ref{MnDopedFieldHist} which show no field history dependence.
For materials with low to moderate pinning field oscillations will either leave the VL unchanged or give rise to a further ordering, as seen for example in YNi$_2$B$_2$C~\cite{Levett:2002ba}.
In contrast, in materials with strong pinning the least disordered VL is in our experience obtained by a field cooling, and any subsequent fields changes will lead to additional disordering.
A final argument against strong pinning comes from the transition kinetics summarized in Fig.~\ref{MnDopedKineticsActivation}.
This shows that it is still possible to induce vortex motion and drive the VL from the metastable to the equilibrium state by small amplitude field oscillations.
At the qualitative level Mn-doped {\MgB} again shows the same behavior as the pure compound, although there are quantitative differences in the saturation value $\fms (n \rightarrow \infty)$.

%%%%% CONCLUSION %%%%%
\section{Conclusions}
We have studied the VL in {\MgB} that has been doped with either manganese or carbon.  
While only minor changes to the phase diagram were observed in the Mn-doped case, only a single VL phase is observed for C-doping within the range of fields and temperatures explored.
This is consistent with the current understanding of superconductivity in {\MgB}.
The VL metastability previously observed in pure {\MgB} is also seen for the Mn-doped case despite a significant reduction of the longitudinal correlation length.

%%%%% BACK MATTER %%%%%
%\section{Data Availability Statement}
%The SANS data for experiment carried out at Institut Laue-Langevin can be found at https://doi.ill.fr/10.5291/ILL-DATA.5-42-420.

\ack{We would like to thank J.~Barker for help with the SANS experiments as NCNR and T.~Williams for assistance collecting the X-ray Laue data.
%\section{Funding information}
Funding for the neutron scattering was provided by the US Department of Energy, Office of Basic Energy Sciences, under Award No. DE-SC0005051. 
S.M. was funded by the Gordon and Betty Moore Foundations EPiQS Initiative through Grant GBMF4411.
The work at Ames was supported by the US Department of Energy, Office of Science, Basic Energy Sciences, Materials Science and Engineering Divison.
Ames Laboratory is operated for the US Department of Energy by Iowa State University under contract No. DE-AC02-07CH11358.
We acknowledge the support of the National Institute of Standards and Technology, U.S. Department of Commerce, in providing the neutron research facilities used in this work.
A portion of this research used resources at the High Flux Isotope Reactor and at the Spallation Neutron Source X-ray laboratory, DOE Office of Science User Facilities operated by the Oak Ridge National Laboratory.}

%%%%% BIBLIOGRAPHY %%%%%
\newpage
\referencelist[MgB2Doped]

%%%%% FIGURES %%%%%
\newpage

\onecolumn

\begin{figure}
    \includegraphics{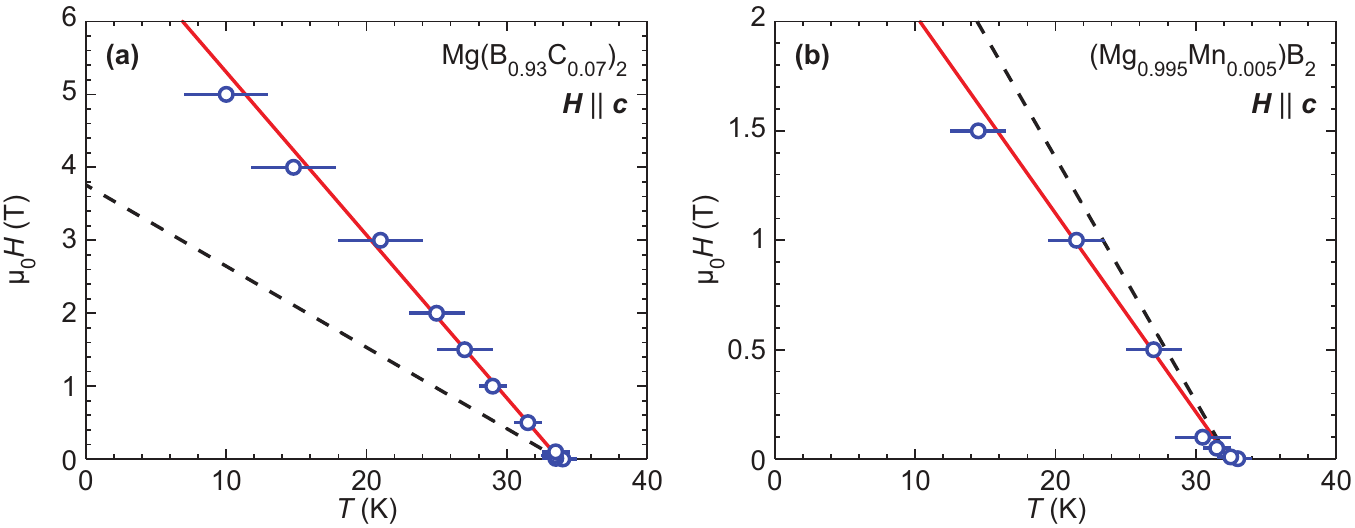}
    \caption{
        Temperature dependence of the upper critical field obtained from magnetization measurements for (a) {\MgBCx} and (b) {\MgMnBx}.
        Solid lines are linear fits to $\Hcii(T)$ close to $\Tc$.
        The dashed lines shows $\Hcii(T)$ for pure Mg$^{11}$B$_2$~\cite{Das:2012cf}, but shifted along the temperature axis to vanish at $\Tc$ for the doped samples}.
    \label{Hc2}
\end{figure}

\begin{figure}
    \includegraphics[width=\columnwidth]{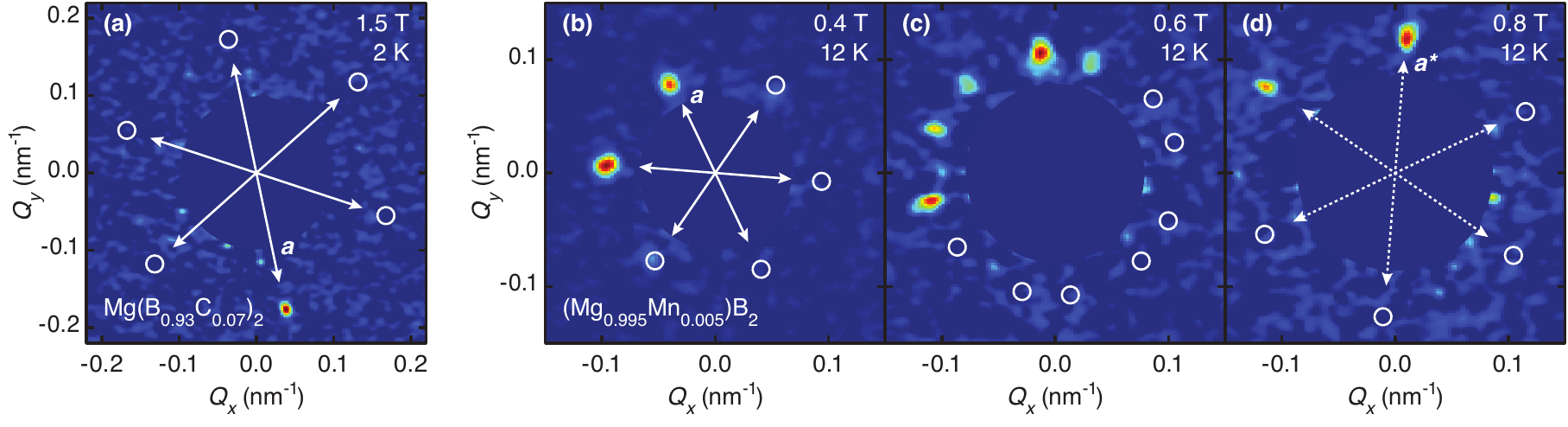}
    \caption{
        Diffraction patterns illustrating the observed VL phases.
        (a) F phase in {\MgBCx}.
        (b) F, (c) L and (d) I phases in {\MgMnBx}.
        Background scattering is subtracted and a portion of reciprocal space close to $Q = 0$ is masked off.
        Crystalline axes in the hexagonal basal plane are indicated by arrows, and the positions of symmetry related VL peaks that were not brought into the Bragg condition are indicated by circles.}
    \label{DifPats}
\end{figure}

\begin{figure}
    \includegraphics[width=\columnwidth]{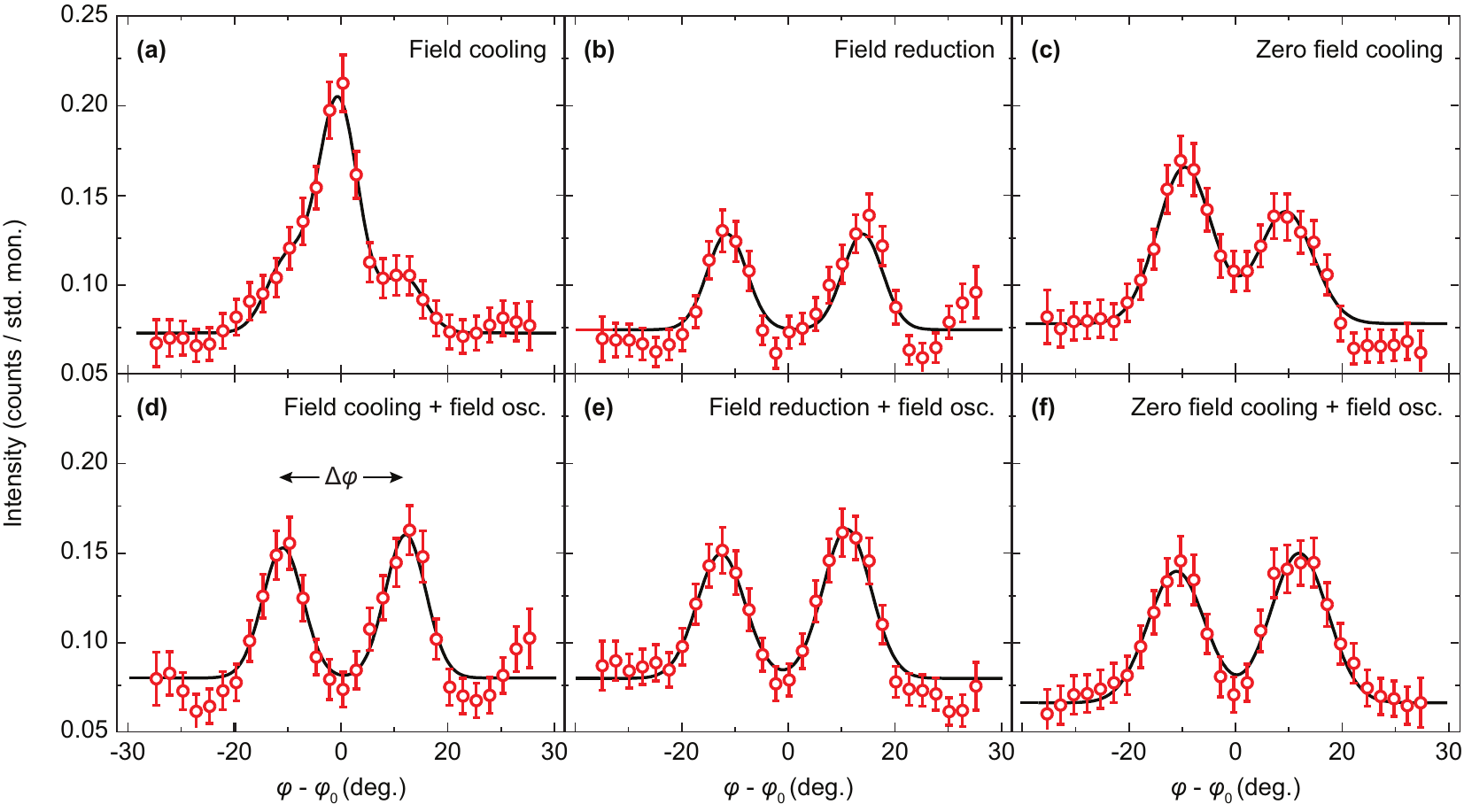}
    \caption{
        Azimuthal VL intensity distribution for {\MgMnBx} at $0.45$~T and $2.1$~K, following different field and temperature histories.
        Top panels correspond to (a) a field cooling, (b) a field reduction from 4~T, and (c) a zero field cooling.
        In the bottom panels (d)-(f) the above field histories were followed by a damped field oscillation.
        Solid lines are multi-Gaussian fits.
        The VL Bragg peak splitting $\Delta \varphi$ is indicated in panel (d).}
    \label{MnDopedFieldHist}
\end{figure}

\begin{figure}
    \includegraphics{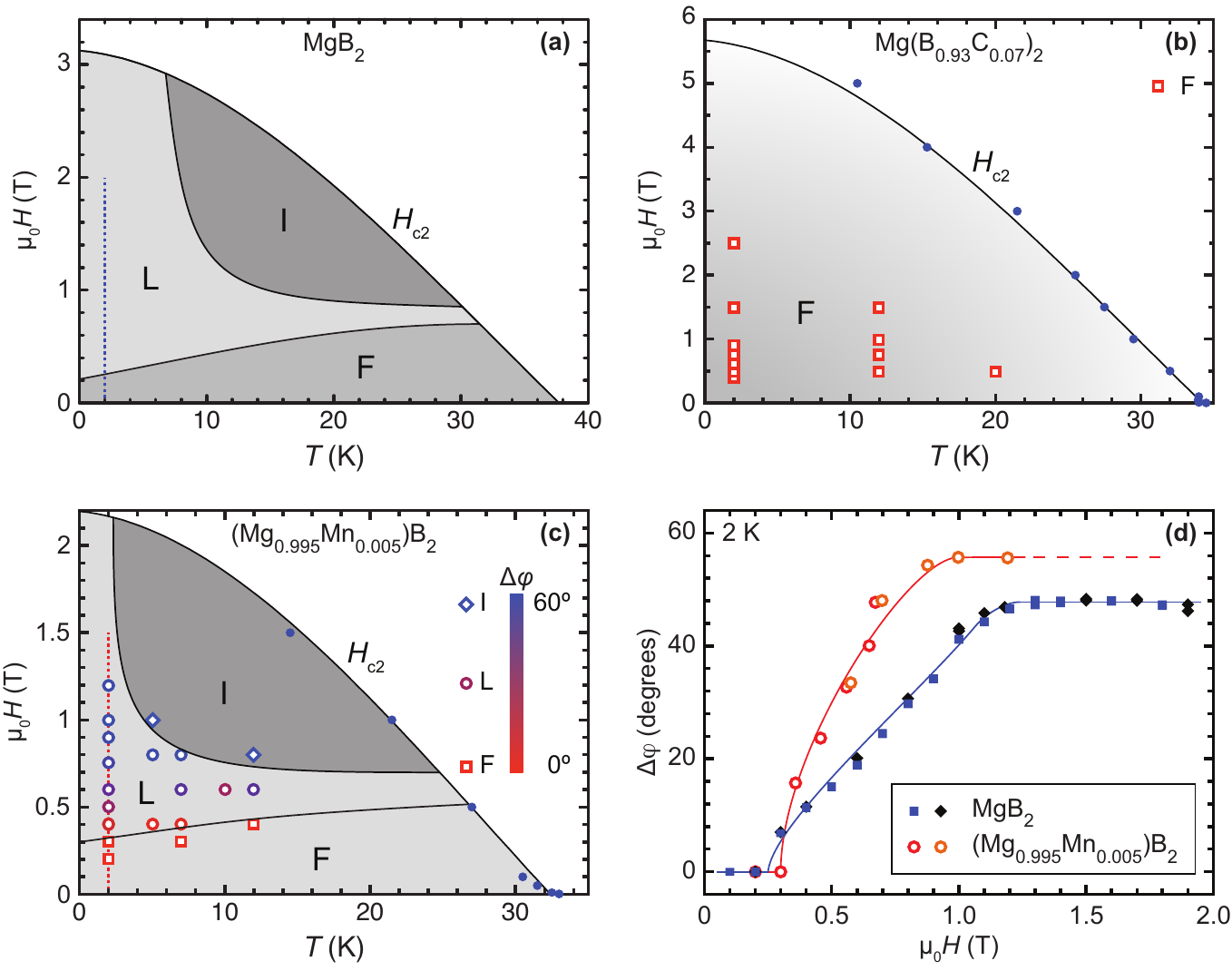}
    \caption{
        VL phase diagrams:
        (a) {\MgB}~\cite{Das:2012cf}, (b) {\MgBCx}, and (c) {\MgMnBx}.
        %Upper critical fields in panels (b) and (c) are based one the data in Fig.~\ref{Hc2} and extrapolated to $T = 0$.
        In panels (b) and (c) the upper critical fields are based one the data in Fig.~\ref{Hc2} and extrapolated to $T = 0$, and vertical dashed lines indicate the field-sweeps in panel (c).
        In (c) the color of the data points reflects the VL Bragg peak splitting $\Delta \varphi$, with $0^{\circ}$ and $60^{\circ}$ corresponding to the F and I phases respectively.
        (d) Comparison of the field dependence of the VL Bragg peak splitting at 2~K for pure~\cite{Das:2012cf} and Mn-doped {\MgB}.
        Errors on the peak splitting do not exceed the size of the symbols.
        Blue/red and black/orange symbols correspond to increasing and decreasing fields respectively.
        All lines are guides to the eye.}
    \label{VLPhaseDiagram}
\end{figure}

\begin{figure}
    \includegraphics[width=\columnwidth]{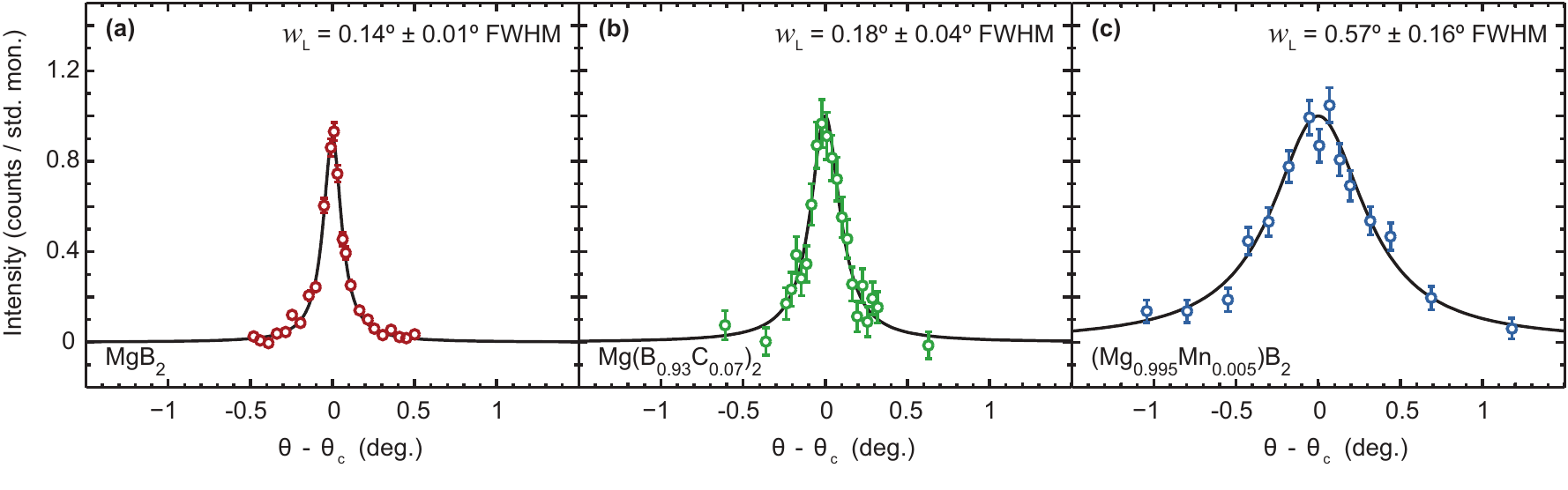}
    \caption{
        VL rocking curves for (a) {\MgB}, (b) {\MgBCx}, and (c) {\MgMnBx}.
        All measurements were performed at 0.5~T and 2~K.
        Solid lines are Voigt fits to the data as discussed in the text, with Lorentzian widths indicated in the individual panels.}
    \label{RockingCurves}
\end{figure}

\begin{figure}
    \includegraphics{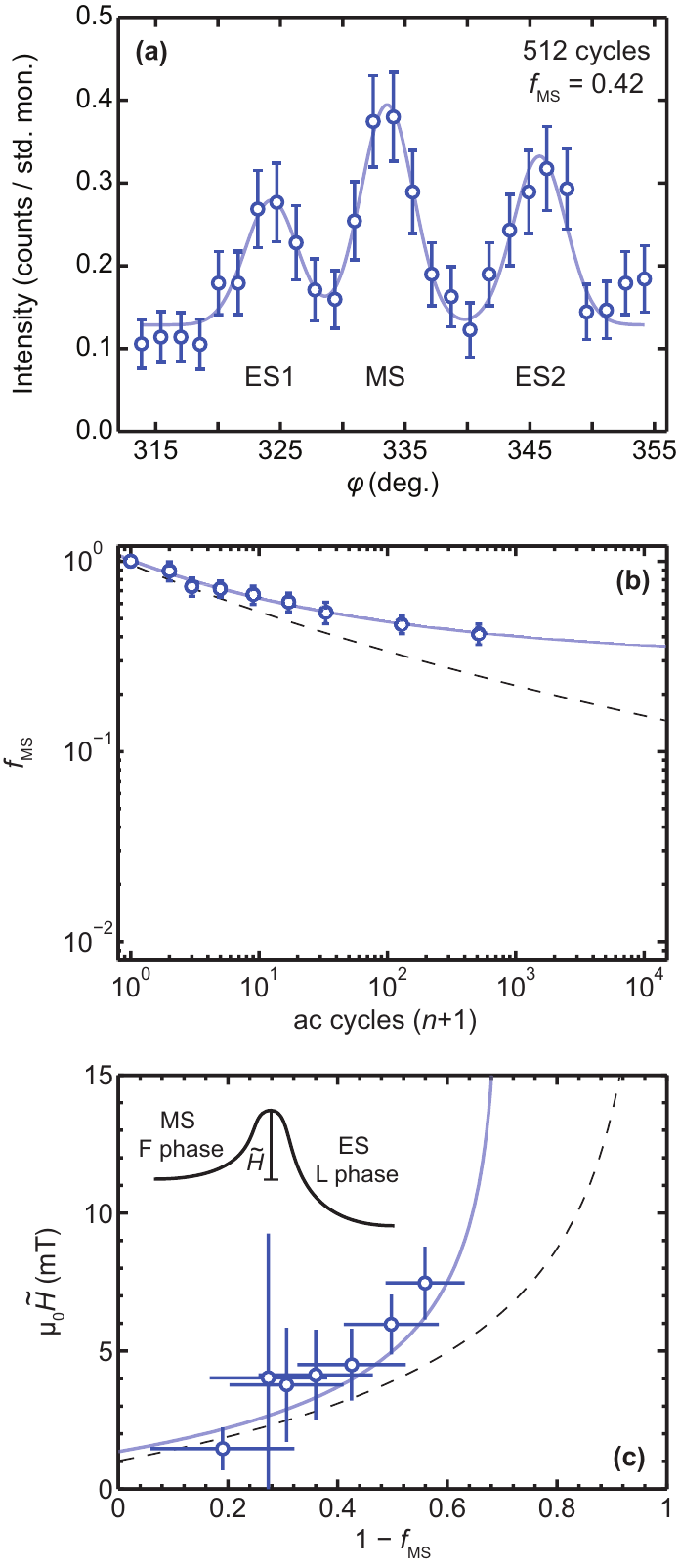}
    \caption{
	    VL transtion kinetics for {\MgMnBx} for a 0.93~mT ac field amplitude with a frequency of 250~Hz.
	    (a) Azimuthal intensity distribution after $512$ field cycles fitted to a three Gaussians of the same width.
	    (b) Metastable VL volume fraction vs the number of applied ac field cycles.
	    The cycle count is offset by one such that $\fms(0)$ may be included, and the solid line is a fit to Eqn.~(\ref{fMSfit}).
	    (c) Activation field determined from $\fms(n)$ using Eqn.~(\ref{Hact}).
	    The solid line is calculated using the parameters obtained from the fit in (b).
	    Dashed lines in (b) and (c) are for pure {\MgB} and the same ac field amplitude and frequency~\cite{Louden:2019bq}.}
	\label{MnDopedKineticsActivation}
\end{figure}

%%%%% SYNOPSIS WILL APPEAR HERE %%%%%
\newpage

\end{document}